\documentclass{emulateapj}
\newcommand{\lsim}{\mbox{$_<\atop^{\sim}$}}
\newcommand{\gsim}{\mbox{$_>\atop^{\sim}$}}

\newcommand{\mum}{$\,\mu$m}

\newcommand{\arcs}{$^{\prime\prime}$}

\newcommand{\chandra}{\textsl{Chandra}}
\newcommand{\spitzer}{\textsl{Spitzer}}
\newcommand{\hst}{\textsl{HST}}
\newcommand{\hz}{\textsc{hyper-z}}
\slugcomment{Accepted for publication in the Astrophysical Journal}

\shorttitle{Stellar and black-hole masses in SCUBA galaxies}
\shortauthors{Borys et al.}

\begin{document}

\title{The Relationship Between Stellar and Black-Hole Mass in 
Submillimeter Galaxies}

\author{
C.\ Borys,$\!$\altaffilmark{1}
Ian Smail,$\!$\altaffilmark{2}
S.\,C.\ Chapman,$\!$\altaffilmark{1}
A.\,W.\ Blain,$\!$\altaffilmark{1}
D.\,M.\ Alexander,$\!$\altaffilmark{3}
R.\,J.\ Ivison$\!$\altaffilmark{4,5}
}
\email{borys@caltech.edu}
\altaffiltext{1}{Caltech, 1200 E California Blvd., Pasadena, CA 91125}
\altaffiltext{2}{Institute for Computational Cosmology, Durham University, South Road, Durham DH1 3LE, UK}
\altaffiltext{3}{Institute of Astronomy, University of Cambridge, Madingley Road, Cambridge CB3 0HA, UK}
\altaffiltext{4}{Astronomy Technology Centre, Royal Observatory, Blackford Hill, Edinburgh EH9 3HJ, UK}
\altaffiltext{5}{Institute for Astronomy, University of Edinburgh, Blackford Hill, Edinburgh EH9 3HJ, UK}

\begin{abstract} 
We analyze deep X-ray, optical and mid-infrared \spitzer\
observations of the CDF-N/GOODS-N region to study a sample of 
13 submillimeter-detected galaxies with spectroscopic 
redshifts (median $z=2.2$).    These galaxies are among the most
active and massive at this epoch.
We find  evidence for 
a power-law correlation between the estimated stellar and  X-ray 
luminosity, implying that masses of the black holes may be related to
the stellar masses of their host galaxies. 
We derive the rest-frame 
UV--near-infrared spectral energy distributions for these galaxies,
believed to be young spheroids, and fit them with model templates.  
Using the rest-frame near-infrared
luminosities, which are relatively insensitive to uncertainties in
stellar ages and 
reddening in these young dusty galaxies, and theoretical mass-to-light
ratios, we can estimate their stellar masses. 
Although the submillimeter
emission implies that these galaxies are undergoing an epoch of intense
star-formation, the \spitzer\ data reveal a 
massive stellar population already in place.
We find that our submillimeter galaxies have a median stellar
mass of $\sim 10^{11}$\,M$_\odot$, which is
roughly ten times more massive than typical UV-selected 
star-forming systems at similar redshifts.
These stellar masses are then compared to previously published
black hole mass estimates derived 
from the X-ray luminosities under the assumption of Eddington-limit accretion.
We find that the black hole masses for our high-redshift sample 
are approximately 1--2 orders of
magnitude smaller than galaxies of comparable stellar mass in the local  
Universe. Although our estimates of black hole masses will
increase if the 
accretion is sub-Eddington, and our stellar masses will decrease if we assume
a much younger stellar population or a different initial mass function, 
we find that only through a combination of effects is it
possible to shift the high redshift galaxies such 
that they lie on the local relation.
This suggests that the black holes need to grow 
substantially between $z=2.2$ and the present-day, with much of the black 
hole growth occurring {\it after} the current obscured, far-infrared luminous 
phase of activity which is likely associated with the formation of the 
spheroid.  This 
interpretation supports a scenario where submillimeter galaxies pass
through a subsequent accretion-dominated phase, where they would 
appear as optically bright quasars.
\end{abstract}


\section{Introduction}
Studies of galaxies in the local Universe have demonstrated an apparent 
relation between the stellar mass of a spheroid and that of the
 super-massive black hole (SMBH) which lies at its heart. 
The 
\mbox{M$_{\star}$--M$_{\rm BH}$} relation for spheroids, with 
the mass of the SMBH typically $\sim 0.1$\% of the spheroid mass
\citep[e.g.][]{1995ARA&A..33..581K, 1998AJ....115.2285M,2000ApJ...543L...5G},
suggests that the growth of the black hole and the surrounding 
galaxy are related. This provides an important clue to understanding 
galaxy formation, as SMBHs have
been suggested as a highly efficient mechanism for regulating star formation 
in galaxies 
\citep[e.g.][]{2002ApJ...570..114C,2004ApJ...600..580G,2005Natur.433..604D}.
Such {\it feedback} is an essential part of current galaxy formation
models and provides a physically-motivated explanation for the form of the 
galaxy luminosity function 
\citep{2003ApJ...599...38B,2005MNRAS.356.1191B}.  
However, observational evidence for feedback, in particular driven by 
activity from SMBHs, is sparse, especially for the youngest and most massive 
galaxies -- for which it is expected to be most important 
\citep{2003ApJ...599...38B}.

Modeling AGN-driven feedback is more difficult than the traditional 
star-formation feedback recipes. In part this is because we still lack a 
clear model for the early growth of the SMBH, where BH--BH mergers and 
super-Eddington accretion may both play a part.  There is thus an urgent 
need for observations of the earliest phases of the evolution of spheroids 
and SMBHs to guide development of theoretical models.

Studies of the submillimeter galaxy (SMG) population 
\citep[e.g.][]{1997ApJ...490L...5S,2000AJ....119.2092B,2002MNRAS.337....1I,
2002MNRAS.331..495S,
2003ApJ...587...41W,2004MNRAS.355..485B,2004ApJ...613..655W,
2004MNRAS.354..779G} 
suggest that luminous far-infrared 
galaxies are likely to be associated with an early-phase in the formation of 
massive galaxies. The intensity of their starbursts, the resulting high 
metallicity, along with their large dynamical masses, high gas fractions 
and strong clustering 
\citep{2002MNRAS.337....1I,gravy,2004ApJ...617...64S,2004ApJ...611..725B}
are all suggestive of a close link to the formation phase of the most massive 
spheroids \citep{2004ApJ...616...71S}.  If these are indeed
young spheroids, what can we learn about their black holes?

Deep X-ray observations provide a sensitive probe of AGN activity in
submillimeter galaxies (SMG) and can be used to place SMBH mass
constraints, even in the presence of heavy gas and dust obscuration.
Exploiting the unique sensitivity of the 2-Ms exposure of the
\chandra\ Deep Field North \citep[\mbox{CDF-N};][]{2003AJ....125..383A}, 
\citet{alexandera,alexanderb} study a small sample of
{\it spectroscopically  identified} SMGs
and find that most host
weak AGN with modest absorption-corrected X-ray luminosities.
This suggest two things: 1) the AGNs do not typically
dominate the bolometric emission from SMGs, and 2) the masses of the
SMBHs are likely to be $\lsim10^8$M$_\odot$ (Eddington-limited SMBH
masses are typically $\sim 10^7$M$_\odot$). The large 
AGN fraction in the SMG population ($\ga$\,40\% for sources with
$f_{\rm 850\mu m}\gsim4$\,mJy) further indicates that
the SMBHs are growing almost continuously throughout these vigorous,
but obscured, 
star-formation episodes. The stark contrast between the AGN fractions
in the SMG and coeval UV-bright galaxy populations ($\sim$\,3\%;
\citet{steidel04}) suggest that this joint black hole--stellar growth
activity is most strongly connected with the SMG population
\citep{alexandera}.

In this paper we compare estimates for the  stellar and SMBH masses 
for a sample of 13 high-redshift proto-spheroids (identified with
far-infrared luminous galaxies detected in the submillimeter 
and radio wavebands) in the 
\mbox{GOODS-N} region.  These all have precise spectroscopically
 identified redshifts and have hard X-ray emission dominated by an AGN.
The stellar mass estimates are derived from optical, near-infrared and 
most importantly the mid-infrared observations of this region by the 
{\it Spitzer Space Telescope} taken as part of the GOODS survey 
\citep{dickinson}, while the SMBH mass estimates are taken from
\citet{alexandera,alexanderb} based on the 2-Ms \chandra\
observations. Together, these data
allow us to investigate the origin of the 
\mbox{M$_{\star}$--M$_{\rm BH}$} relation for young spheroids in the 
early Universe. Throughout, we adopt a cosmology with 
H$_0=65$km\,s$^{-1}$\,Mpc$^{-1}$, 
$\Omega_{\rm M}=1/3$, and $\Omega_\Lambda=2/3$.  Unless otherwise
stated, masses and luminosities are given in solar units.

\section{Sample Selection and Data Reduction}
For local galaxies, spatially resolved kinematics 
\citep[e.g.][]{1998AJ....115.2285M} or 
reverberation mapping \citep[e.g.][]{1999ApJ...526..579W,2000ApJ...533..631K} 
 are useful techniques for weighing the 
SMBHs at the center of galaxies and relating these to the stellar
masses
of their spheroids.  However,
neither of these techniques can
currently
be applied to SMGs at high redshifts and so by necessity we have to
adopt much cruder methods to estimate both the black hole and spheroid
masses.  In this study we use the emission arising from accretion onto
the SMBH
as a tracer of the black hole mass and the rest-frame
near-infrared luminosity of galaxy as
an indicator of its stellar mass.  

Unfortunately, the rest-frame UV and optical spectra of the SMGs 
provides an ambiguous tracer of the nuclear activity in 
these galaxies
\citep{chapmanrsmg,2004ApJ...617...64S}. The difficulty
of merely identifying an AGN
from these observations,
let alone reliably estimating their luminosities is likely caused
by heavy obscuration towards the SMBH or faint broad lines from  intrinsically
weak AGN being lost against the continuum of the galaxy.
Conversely the
weakness of the AGN and their obscuration makes the use of the
rest-frame near-infrared luminosity a much more reliable measure of the
stellar mass of the galaxies.

In contrast,
hard X-ray emission from the accretion disk emerges relatively
unattenuated from all but the most obscured regions, hence the
X-ray emission from SMGs may provide the most reliable probe of
their SMBH masses. 
However, as demonstrated by \citet{alexandera}, 
only the {\it Chandra} 2-Ms observation 
of the \mbox{CDF-N} region \citep{2003AJ....125..383A}
reaches the necessary depth to provide constraints on the X-ray luminosities
of typical SMGs. For this reason our analysis focuses on the
sample of SMGs with reliable spectroscopic redshifts
from \citet{2003Natur.422..695C,chapmanrsmg} which lie within this field.
We stress that 
precise redshifts are essential for our analysis.  
Not only are the mass estimates dependent
on a well-constrained luminosity distance, but 
more critically there are strong degeneracies in redshift and age
in the  fitting of their spectral energy distributions (SED)
described in \S3.

\subsection{X-ray constraints on SMGs}
Using the sample of spectroscopically identified, 
radio-detected SMGs in the CDF-N from
\citet{chapmanrsmg}, obscuration-corrected X-ray luminosities were 
derived for 15 galaxies
and SMBH masses estimated assuming Eddington accretion by
\citet{alexandera,alexanderb}.  
We discard two of these galaxies from our analysis:
J123716.01+620323.3 falls outside the mid-infrared coverage,
and the optical/mid-infrared photometry of J123632.61+620800.1 is contaminated 
by a diffraction spike from a bright ($K_s\sim14$) object 
only 11\arcs\ away.  These conditions prevent us from estimating the 
stellar masses needed in \S3.2, 
hence we concentrate on 13 SMGs with optical,
near-infrared and mid-infrared photometry or limits and
reliable spectroscopic redshifts. These galaxies range in redshift 
from $z=0.5$--2.9, with a median comparable to that of the entire SMG 
population ($z=2.2$).  

The X-ray spectral analysis in \citet{alexanderb} shows 
that the bulk of the 
hard X-ray emission from the SMGs in our sample is likely to be
due to accretion onto the
SMBH, with star formation contributing only at the softest energies,
$\sim 2$\,keV in the rest-frame.
The  luminosity from the X-ray waveband 
is only a few percent of the bolometric luminosity from the
SMBH, and \citet{alexanderb} adopted a
bolometric correction (BC$_X$) of $6^{+12}_{-4}$\%
consistent with other studies. They also
employed Eddington arguments to show that 
$\log{(M_{\rm BH})}=\log(L_{\rm bol}/\eta) - 38.06$, 
where the bolometric luminosity is in units of erg\,s$^{-1}$. 
The factor $\eta$ is unity for 
Eddington-limited accretion, and $\eta <1$ for
sub-Eddington.   As our bench-mark we adopt Eddington-limited
accretion and use this to 
investigate the relation between black hole and stellar
masses for our sample.

\subsection{Spitzer observations}
In addition to the X-ray constraints on the luminosities and hence masses
of the SMBHs, our analysis also requires estimates of the stellar masses
of the host galaxies.  A particularly simple and relatively robust way of 
deriving the mass of a stellar population is to use the rest-frame 
near-infrared luminosity and a theoretical mass to light ratio. For all 
but the youngest stellar populations, the rest-frame near-infrared emission 
is dominated by light from main sequence stars and hence, for example, the 
rest-frame $K$-band luminosity changes by only a factor of two between 
stellar populations with ages of 100\,Myr and 1\,Gyr 
\citep{1999ApJS..123....3L}. The 
rest-frame $V$-band luminosities of the same populations on the other hand
vary by a factor of 
ten. In addition to its relative 
insensitivity to age variations, the rest-frame $K$-band is over
ten times less sensitive to reddening than $V$-band, which is important
to consider for these dusty SMGs.

At the redshift of our targets, the rest frame $K$-band is shifted towards the
mid-infrared and hence {\it Spitzer} observations are required to estimate
the stellar mass. For this reason, we obtained the IRAC images of 
GOODS-N from the 
\spitzer\ archive (V.1 20041027 and V.1 20050505 enhanced data products).
We used \textsc{Sextractor} \citep{1996A&AS..117..393B} to measure the 
photometry and uncertainties
in these for each of the galaxies on the IRAC imaging.  By
constructing curves of growth for
several isolated objects, we determined that 6\arcs\ diameter apertures were 
necessary to
measure the total IRAC flux from our targets.  Since source confusion 
is an issue for these
extremely deep exposures, we used 4\arcs\ diameter 
apertures and then applied a small
aperture correction, determined to be 0.22, 0.26, 0.33, and 0.30 mag for the
3.6, 4.5, 5.8, and 8.0\mum\ bands respectively.  Even so, we were unable 
to adequately constrain the 3.6\mum\ flux for three sources due to 
confusion noise from nearby sources.
We provide the IRAC 
photometry for all 
13 sources in Table~\ref{tab:phot-spitzer}.
At 5.8$\mu$m, which corresponds to a rest-frame wavelength of $\sim 2\mu$m,
the ten high-redshift SMGs at $z>1.8$ have a bright mean magnitude of
$20.1\pm 0.8$, with the
entire sample  spanning under an order of magnitude in flux.
This is in contrast with the $R$-band fluxes, which cover well over an order
of magnitude.
The modest dispersion in the mid-infrared fluxes of 
the SMGs already suggests that we are dealing with a relatively
homogeneous population of luminous, high redshift galaxies.

\begin{deluxetable*}{lcccc}
\tablecaption{\spitzer\ IRAC photometry}
\tablehead{
\colhead{ID} & 
\colhead{3.6$\mu$m} & \colhead{4.5$\mu$m} & \colhead{5.8$\mu$m} &\colhead{8.0$\mu$m}
}
\startdata
J123636.75+621156.1                    & $20.61\pm0.08$ & $20.96\pm0.08$ & $21.18\pm0.22$ & $21.44\pm0.14$\\
J123721.87+621035.3                    & $19.35\pm0.04$ & $19.66\pm0.04$ & $19.96\pm0.11$ & $20.12\pm0.05$\\
J123629.13+621045.8                    & $19.06\pm0.04$ & $19.13\pm0.04$ & $19.37\pm0.10$ & $19.59\pm0.06$\\
J123555.14+620901.7                    & $19.91\pm0.10$ & $19.51\pm0.08$ & $19.31\pm0.17$ & $19.29\pm0.09$\\
J123711.98+621325.7                    & $21.56\pm0.10$ & $21.33\pm0.08$ & $21.08\pm0.15$ & $21.52\pm0.08$\\
J123635.59+621424.1                    & $19.45\pm0.04$ & $18.98\pm0.03$ & $18.46\pm0.05$ & $18.03\pm0.02$\\
J123549.44+621536.8\tablenotemark{a,b} & \nodata        & $19.78\pm0.30$ & $19.61\pm0.30$ & $19.63\pm0.30$\\
J123606.72+621550.7                    & $21.52\pm0.17$ & $21.18\pm0.11$ & $20.85\pm0.25$ & $20.68\pm0.11$\\
J123622.65+621629.7                    & $20.74\pm0.08$ & $20.38\pm0.05$ & $20.19\pm0.11$ & $20.69\pm0.06$\\
J123707.21+621408.1                    & $20.78\pm0.10$ & $20.44\pm0.08$ & $20.18\pm0.14$ & $20.55\pm0.05$\\
J123606.85+621021.4                    & $20.37\pm0.08$ & $20.25\pm0.06$ & $20.19\pm0.15$ & $20.61\pm0.10$\\
J123616.15+621513.7\tablenotemark{b}   & \nodata        & $20.76\pm0.08$ & $20.36\pm0.15$ & $20.05\pm0.07$\\
J123712.05+621212.3\tablenotemark{b}   & \nodata        & $21.06\pm0.07$ & $20.78\pm0.13$ & $20.83\pm0.05$\\
\enddata
\tablecomments{All magnitudes are in the AB system. 
Values represent 6\arcs\ diameter apertures, derived by applying a 
correction to the measured 4\arcs\ aperture magnitudes.}
\tablenotetext{a}{This sources lies on the edge of the image,
  although is clearly detected.
In these cases we set the error to 0.3\,mag to reflect the
uncertainty in our aperture fluxes.}
\tablenotetext{b}{The 3.6$\mu$m images for these sources were confused, and {\sc sextractor} could not converge on a solution.  Since the other photometry is sufficient to constrain the shape of the SED, we choose not to use the data from this band.}
\label{tab:phot-spitzer}
\end{deluxetable*}

\subsection{Spectral Energy Distributions}
To improve the constraints on the form of the SEDs in SMGs we also make use of 
rest-frame UV/optical photometry to extend the wavelength coverage and so 
estimate the likely reddening and ages of the stellar populations. 
For this we use the $UBVIR{z^\prime}$ photometry presented in
\citet{2004AJ....127..180C}. These data were supplemented with  
$J$- and $K_s$-band near-infrared imaging from the WIRC2 camera at 
Palomar (see \citet{bundy} and
\citet{2004ApJ...616...71S} for a description of these data).
Combined, these ground-based 
data cover more bands than the GOODS-N 
{\it Hubble Space Telescope} (\hst) 
ACS imaging of this field \citep{2004ApJ...600L..93G}.
Although the extra bands provided by the ground based imaging is
essential for the 
SED template fitting (see \S\ref{sec:stellarmass}), the high-resolution 
imaging available from {\it HST} 
is valuable for studying the light profiles of these galaxies.
\citet{boryshighres} show that the rest-frame $V$-band
light distributions of a sample of 27 SMGs (5 of which are in the sample
presented here) follow a $r^{1/4}$ power law on average,
consistent with the spheroidal-nature of these systems.
A summary of the
optical and near-IR photometry is presented in
Table~\ref{tab:phot-opt}.
We adopt 3\arcs\ and 4\arcs\ diameter apertures to measure the fluxes
from the optical and near-infrared imaging as these provide optimal
signal-to-noise and a near-total estimate of the 
magnitudes for the SMGs.

\begin{deluxetable*}{lllllllll}
\tabletypesize{\scriptsize}
\tablecaption{Ground-based optical and near-IR photometry}
\tablewidth{0pt}
\tablehead{
\colhead{ID} & 
\colhead{$U$} & \colhead{$B$} & \colhead{$V$} & \colhead{$R$} & 
\colhead{$I$} & \colhead{$z^\prime$} & \colhead{$J$} & 
\colhead{$K_s$\tablenotemark{a}} 
}
\startdata
J123636.75+621156.1 &$24.28\pm0.08$ &$23.43\pm0.05$ &$23.09\pm0.03$ &$22.39\pm0.02$ &$21.84\pm0.02$ &$21.61\pm0.03$ &$21.12\pm0.04$ &$20.58\pm0.04$ \\
J123721.87+621035.3 &$24.19\pm0.08$ &$24.00\pm0.07$ &$23.52\pm0.04$ &$22.81\pm0.02$ &$21.84\pm0.03$ &$21.35\pm0.02$ &\nodata        &$19.84\pm0.02$ \\
J123629.13+621045.8 &$26.11\pm0.25$ &$25.70\pm0.15$ &$24.98\pm0.09$ &$24.05\pm0.04$ &$22.89\pm0.04$ &$22.36\pm0.04$ &$21.39\pm0.05$ &$19.93\pm0.02$ \\
J123555.14+620901.7 &$24.38\pm0.09$ &$24.24\pm0.08$ &$24.16\pm0.05$ &$23.80\pm0.04$ &$23.45\pm0.06$ &$23.16\pm0.06$ &$21.99\pm0.09$ &$20.92\pm0.05$ \\
J123711.98+621325.7 &$26.12\pm0.26$ &$25.75\pm0.14$ &$25.71\pm0.16$ &$25.66\pm0.13$ &$24.99\pm0.18$ &$25.06\pm0.24$ &$>24.0       $ &$>23.1       $ \\
J123635.59+621424.1 &$23.94\pm0.07$ &$24.06\pm0.07$ &$24.00\pm0.05$ &$23.79\pm0.04$ &$23.36\pm0.06$ &$23.11\pm0.06$ &$21.99\pm0.09$ &$20.74\pm0.04$ \\
J123549.44+621536.8 &$24.61\pm0.10$ &$23.93\pm0.07$ &$23.60\pm0.04$ &$23.27\pm0.03$ &$23.01\pm0.04$ &$22.84\pm0.05$ &$21.77\pm0.07$ &$21.49\pm0.18$\tablenotemark{a}\\
J123606.72+621550.7 &$23.75\pm0.06$ &$23.32\pm0.06$ &$23.29\pm0.03$ &$23.26\pm0.03$ &$23.09\pm0.05$ &$23.06\pm0.06$ &$>24.0       $ &$>22.8       $\tablenotemark{a}\\
J123622.65+621629.7 &$25.62\pm0.17$ &$25.17\pm0.14$ &$25.09\pm0.11$ &$24.85\pm0.08$ &$24.45\pm0.13$ &$24.21\pm0.14$ &$>24.0       $ &$22.25\pm0.15$ \\
J123707.21+621408.1 &$>27.8       $ &$26.77\pm0.27$ &$26.36\pm0.29$ &$25.63\pm0.14$ &$25.51\pm0.30$ &$24.69\pm0.19$ &$>24.0       $ &$21.86\pm0.11$ \\
J123606.85+621021.4 &$26.17\pm0.26$ &$25.25\pm0.14$ &$25.15\pm0.10$ &$24.41\pm0.06$ &$24.21\pm0.10$ &$23.93\pm0.10$ &$22.20\pm0.10$ &$20.98\pm0.05$ \\
J123616.15+621513.7 &$>27.8       $ &$26.59\pm0.24$ &$26.13\pm0.22$ &$25.36\pm0.11$ &$25.03\pm0.18$ &$25.00\pm0.23$ &$>24.0       $ &$22.71\pm0.23$ \\
J123712.05+621212.3 &$>27.8       $ &$>27.6       $ &$>27.6       $ &$>27.4       $ &$>26.4       $ &$>26.2       $ &$>24.0       $ &$23.02\pm0.30$ \\
\enddata
\tablecomments{All magnitudes are in the AB system.  For undetected sources, 
we quote $2.5\sigma$ upper limits. The $J$ and $K_s$ magnitudes are measured 
in 4\arcs\ apertures, while all other bands use 3\arcs.}
\tablenotetext{a}{Two sources that fall outside our $K_s$ coverage are 
in the $HK^\prime$ catalog provided by \citet{2004AJ....127..180C}, so we 
use those values here.}
\label{tab:phot-opt}
\end{deluxetable*}

\begin{figure*}
\plotone{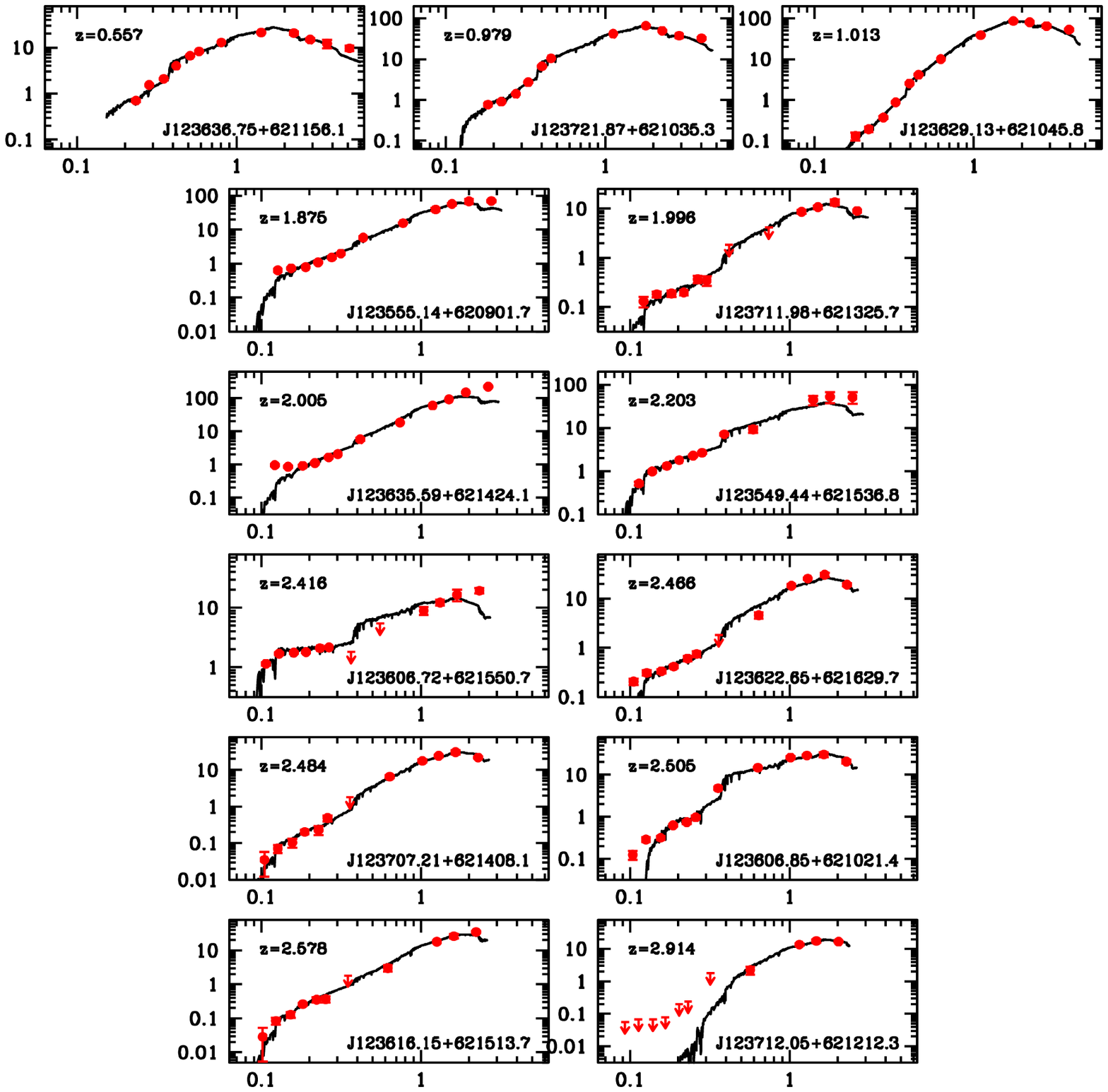}
\caption{The spectral energy distributions of the 13 SMGs with black hole mass estimates 
from \citet{alexanderb}, ordered in increasing redshift.  
The top line shows
the SEDs of the lower redshift SMGs, while the remaining 10 are
all $z>1.8$.
The best-fit template from \hz\ is overlaid on the measured photometry 
(in units of $\mu$Jy) and the 
wavelengths (shown in units of \mum) are corrected to the rest-frame
at the redshift noted. 
}
\label{fig:allseds}
\end{figure*}

We plot the combined optical, near- and mid-infrared photometry
for the 13 SMGs in Fig.~\ref{fig:allseds}.
The SEDs for these galaxies show 
three clear 
features:  a sharp decline at rest-frame wavelength of $\leq 912$\AA, 
resulting from the Lyman$-\alpha$ limit within the galaxy and the foreground
intergalactic medium; a break at around 4000\AA\ arising either from
the spectral breaks found in young or
evolved stellar populations at this approximate
wavelength and finally a peak in the 
emission at $\sim1.6\mu$m associated with the minimum in the H$^-$ opacity 
\citep{1988A&A...193..189J,1999PASP..111..691S} and hence the maximum 
emission from stellar atmospheres.  

Looking at the SEDs in Fig.~\ref{fig:allseds} we conclude that 
there is a modest range in the rest-frame 
UV/optical/near-infrared properties 
of radio-detected SMGs. However, focusing on the ten higher-redshift SMGs, at
$z>1.8$,  of those with full coverage of their
rest-frame UV and optical SEDs, 80\% 
show a detectable discontinuity in their SEDs around  
4000\AA\ in the rest-frame (Fig.~\ref{fig:allseds}).  We
interpret this break as  arising from the Balmer limit in stellar populations
with ages of $\sim 100$'s Myrs.  This interpretation is consistent
with the likely nature of SMGs as luminous far-infrared sources seen during
a phase of intense star formation.  We note that age-related 
differential reddening
may also account for the relative strength of this feature in an even
younger on-going starburst \citep{2000ApJ...529..157P}.

\section{Analysis and Results}
\label{sec:masses}

\subsection{Estimating Stellar Masses}
\label{sec:stellarmass}
Due to the complex mix of extinction expected within SMGs 
\citep{2004ApJ...611..732C}, which may influence different age 
stellar populations to different
degrees, we have adopted a very simple approach in determining our stellar
masses.  We first estimate the rest-frame $K$-band luminosities of
the galaxies, interpolated if necessary from a fit to the SED, and
then combine these with estimates of the mass-to-light ratios
in the $K$-band for a range of plausible ages for the dominant 
stellar population.

To derive the rest-frame $K$-band luminosities
we  exploit the \hz\ package
\citep{2000A&A...363..476B} to fit physically
meaningful spectral templates to the  spectral energy 
distributions of each galaxy and then estimate
their  $K$-band luminosities by interpolation of the best-fit templates.
This approach has the added advantage that the properties of the best-fit
spectral templates may provide some insights into the
properties of the stellar populations within these galaxies.
We restrict the code to fit the SEDs at the spectroscopically
confirmed redshift, and allow it to choose between  
two model SED templates:  one having an instantaneous burst of star-formation, 
or one with a constant star formation (CSF) history. 
These represent the two extremes of the likely star formation 
histories for these young galaxies \citep{2003MNRAS.342.1185S}
and provide an adequate variety to guarantee a reliable
estimate of the $K$-band luminosity for those galaxies whose
SEDs are dominated by star-light.
In our fits, the reddening ($A_V$) was free to vary, and the ages at the 
galaxy's redshift are constrained to be physical within our adopted
cosmology.  These provide additional degrees of freedom for the fitting
and may give some indication of the properties of the composite
stellar population in these systems.

For each galaxy, \hz\ returns 
the absolute magnitude in the rest-frame $K$-band, as well
as the  age, reddening and star formation history of the 
best-fit template.
Although the  M$_K$ values are not corrected for it, the reddening (which
is $\sim0.1$A$_V$ based on the extinction law presented 
in \citet{2000ApJ...533..682C}) has a negligible effect on our stellar
mass estimates, particularly compared with other uncertainties in our
analysis.
The typical fit to a SED in our sample produces a reasonable $\chi^2$, 
with only three minor exceptions:
Both J123635.59+621424.1 and J123606.72+621550.7 are spectroscopically
identified AGN, and demonstrate 
an obvious excess at the reddest wavelengths.  This would
be expected, since the input templates do not account for the 
contribution in the mid-IR of warm dust from an AGN.  J123549.44+621536.8
also has a moderate near-IR excess, but the significance is muted somewhat by
its proximity to the edge of the IRAC map, 
where photometry was more difficult to
estimate.
The starburst 
J123606.85+621021.4 has an
excess in the UV, but this is caused by a contribution from an unrelated blue
foreground $(z=0.4)$ galaxy only 2\arcs\ away.
In all these cases, we feel the best-fit SED template is
still a good measure of the rest-frame $K$-band luminosity and  provides
a good estimate of the underlying stellar mass.  As we show later,
changes to the implied stellar masses of factors of 2 (due to
AGN contamination at restframe $K$-band), would
not significantly alter the conclusions of this work.

To convert these estimates for M$_K$  into
stellar masses we need to determine the light-to-mass ratio,
L$_K$/M for the dominant stellar populations in these
galaxies.  To do this we turn to the 
{\sc Starburst99} stellar population model 
\citep{1999ApJS..123....3L}, which provides estimates
of  M$_K$ for our two extreme scenarios -- either a 
burst or constant star formation history.  We adopt the 
the initial mass function (IMF) 
of \citet{1979ApJS...41..513M}, this is
also the same as that used for the \hz\ templates.  
The IMF has a mass range
of 0.1--125\,M$_\odot$ and assumes solar metallicity.  
Higher metallicities are
often favored  when fitting the SEDs of
Ultraluminous Infrared Galaxies (ULIRGS) in the local 
Universe \citep[e.g.][]{farrah_stis}, 
but we found no appreciable difference in the estimates of L$_K$/M
for a range of likely values.

We can combine the predicted $K$-band luminosities as a function of
time from {\sc  Starburst99} 
with the integrated stellar masses at that time to derive the
effective light-to-mass ratio for the stellar population:
\begin{equation}
{\rm L}_K/{\rm M}= 10^{-0.4(M_K -3.3)} / \Sigma (M_*(t)),
\end{equation}
where 3.3 is the absolute magnitude of the Sun in the 
$K$-band \citep{2000asqu.book.....C}.  We
find that the 
light-to-mass ratio for the CSF case is reasonably 
described by a simple power law, 
\mbox{L$_K$/M$ = 103(\tau$/Myr)$^{-0.48}$} for $\tau>100\,$Myr.
The behavior of L$_K$/M for a  burst model is not as easy to 
parameterize over the likely range of ages of
SMGs, so instead we use tabulated values.  In solar units, we
find that L$_K$/M is [2.5, 2.2, 13.3, 7.6, 4.2, 2.7, 1.9] at 
10, 20, 50, 100, 200, 500, 1000\,Myrs  respectively.
We caution the reader that at ages of $<50$\,Myrs  changes to
the adopted IMF or metallicity can have significant effects
on the L$_K$/M and so the predictions for extremely young
bursts are much more uncertain.

In principle, 
we can constrain the ages of the dominant stellar population 
in each SMG from our
SED modeling. However,  
the correlated nature of the fits and the fact that we are likely
sampling different mixes of stellar populations at different wavelengths
due to differential reddening means that the results are
highly uncertain for individual galaxies. 
To demonstrate this point, we 
note that all of the galaxies can be reasonably fit by {\it either} 
model
star formation history, but with very different ages.  This is 
a direct consequence of the fact that an older galaxy with a 
constant star-formation history has a similar SED to that of a younger 
burst model, a point also noted in
\citet{2001ApJ...562...95S,aliceirac}. 
For this reason
we prefer to use only ensemble averages for the derived parameters.
If we restrict our \hz\ fits to the 
burst model only, we estimate ensemble averages of $250\pm190$\,Myr, 
$A_V=1.7\pm0.2$ and  M$_K=-26.3\pm0.3$.  
Alternatively, restricting our fits to just  the
continuous star-forming model yields  $2000\pm500\,$Myr, 
$A_V=1.7\pm0.3$ and  M$_K=-26.4\pm0.3$.  

For the average age from our fits of the burst model,
 we estimate a typical L$_K$/M=3.7,
while the average for the 
continuous star-forming model 
is L$_K$/M=2.6.  Hence, the average stellar mass from the samples, 
estimated using $M_K$ and
 L$_K$/M, are essentially identical, $\log(M_{\rm burst})=11.3\pm0.4$
and $\log(M_{\rm CSF})=11.5\pm0.4$.  Based on these results,
we choose to simply use  L$_K$/M=3.2 for all galaxies.
Given the degeneracy between the SED fits, we use the average M$_K$ from
the burst and CSF fit for each galaxy, and half the difference between them
as a measure of the uncertainty.
We summarize the results 
of our stellar mass estimates from the SED
fitting in Table~\ref{tab:hyperz}. The mean stellar mass is  
$\log(\rm{M_{\star}})=11.4\pm0.4$ for the sample of SMGs used here.

\begin{deluxetable*}{llllccl}
\tablecaption{Derived optical and X-ray parameters for a sample of 
SMGs in GOODS-N}
\tablehead{\colhead{ID} & 
\colhead{$z$\tablenotemark{a}} & 
\colhead{M$_K$\tablenotemark{b}} &
\colhead{$\log(M_{\star})$\tablenotemark{c}} &
\colhead{$\log(L_\textrm{X})$\tablenotemark{d}} &
\colhead{$\log(M_{\rm BH})$\tablenotemark{d}} &
\colhead{Fit quality\tablenotemark{e}}
}
\startdata
J123636.75+621156.1                  & 0.557 & $-24.10\pm0.06$ & $10.46\pm0.02$ & 42.7 & $6.0^{+0.3}_{-6.0}$ & \\
J123721.87+621035.3                  & 0.979 & $-26.14\pm0.03$ & $11.27\pm0.01$ & 43.7 & $6.8^{+0.5}_{-0.5}$ & \\
J123629.13+621045.8                  & 1.013 & $-26.32\pm0.05$ & $11.34\pm0.02$ & 43.2 & $6.3^{+0.5}_{-0.3}$ & \\
J123555.14+620901.7                  & 1.875 & $-27.28\pm0.37$ & $11.73\pm0.15$ & 44.4 & $7.6^{+0.5}_{-0.5}$ & \\
J123711.98+621325.7                  & 1.996 & $-25.48\pm0.40$ & $11.01\pm0.16$ & 43.6 & $6.8^{+0.5}_{-0.5}$ & \\
J123635.59+621424.1\tablenotemark{f} & 2.005 & $-27.98\pm0.49$ & $12.01\pm0.20$ & 44.0 & $7.1^{+0.5}_{-0.4}$ & Near-IR Excess\\
J123549.44+621536.8\tablenotemark{f} & 2.203 & $-27.44\pm0.32$ & $11.79\pm0.13$ & 44.0 & $7.1^{+0.5}_{-0.5}$ & Near-IR Excess\\
J123606.72+621550.7                  & 2.416 & $-27.57\pm0.30$ & $11.85\pm0.12$ & 43.8 & $7.0^{+0.5}_{-0.4}$ & Near-IR Excess\\\
J123622.65+621629.7\tablenotemark{f} & 2.466 & $-26.34\pm0.58$ & $11.35\pm0.23$ & 44.0 & $7.2^{+0.5}_{-0.5}$ & \\
J123707.21+621408.1\tablenotemark{f} & 2.484 & $-26.40\pm0.19$ & $11.37\pm0.08$ & 43.8 & $6.9^{+0.5}_{-0.4}$ & \\
J123606.85+621021.4\tablenotemark{f} & 2.505 & $-27.00\pm0.37$ & $11.61\pm0.15$ & 43.7 & $6.8^{+0.5}_{-0.5}$ & UV Excess\\
J123616.15+621513.7                  & 2.578 & $-26.60\pm0.12$ & $11.46\pm0.05$ & 43.7 & $6.8^{+0.5}_{-0.5}$ & \\
J123712.05+621212.3                  & 2.914 & $-25.82\pm0.15$ & $11.15\pm0.06$ & 43.4 & $6.6^{+0.5}_{-0.6}$ & \\
\enddata
\tablecomments{}
\tablenotetext{a}{Redshift determined from UV spectroscopy  \citep{chapmanrsmg}.}
\tablenotetext{b}{Absolute $K$-band magnitude derived from \hz\ (uncorrected for reddening).}
\tablenotetext{c}{Stellar mass in units of M$_\odot$ derived assuming L$_K$/M=3.2.}
\tablenotetext{d}{Rest-frame 0.5--8.0\,keV luminosity (in units of erg\,s$^{-1}$) and 
SMBH mass in units of M$_\odot$ from \cite{alexandera} assuming a
bolometric correction of 6\%, with an uncertainty reflecting the
possible range in this correction.}
\tablenotetext{e}{SED fit was deemed good unless otherwise notes.  See \S3.1.}
\tablenotetext{f}{These five SMGs  also have rest-frame optical
  spectroscopy 
from \citet{2004ApJ...617...64S}}
\label{tab:hyperz}
\end{deluxetable*}

\subsection{The M$_{\star}$--M$_{\rm BH}$ relation}
\label{sec:bhmass}
\begin{figure}
\plotone{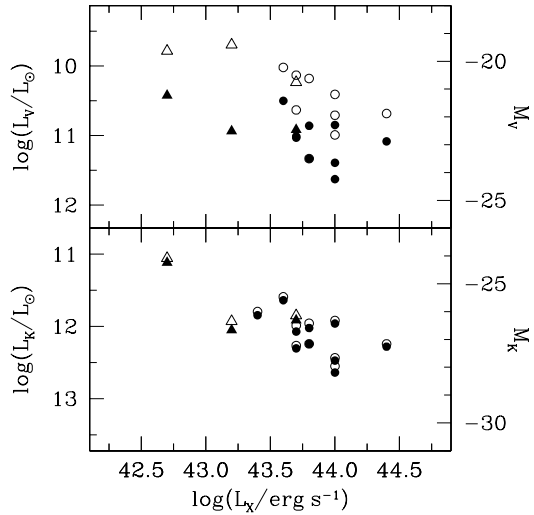}
\caption{We show the correlation between the rest-frame $V$- and $K$-band
luminosities and the rest-frame 
absorption-corrected 2--10\,keV  luminosities of the 13 high-redshift, 
SMGs in our sample. Open symbols represent the 
optical luminosities uncorrected for reddening.  
The best fit value of $A_V$ from \hz\ were used to correct these values for 
reddening, and we show them as the solid points. 
Circles represent the 10 $z>1.8$ subsample, 
while triangles are the three lower redshift sources.
The $K$-band panel 
shows a clear trend for more luminous galaxies to have higher X-ray 
luminosities, as expected 
if they host more massive black holes. A similar trend may be seen
in
the $V$-band panel, however, this is masked by the increased
sensitivity to the reddening correction and the corresponding uncertainties.
In order to keep the scales the same between the two panels, we excluded
J123712.05+621212.3 from the $V$-band plot since it falls
considerably outside the range (with $M_V=-13.5$).}
\label{fig:m-lx}
\end{figure}

Fig.~\ref{fig:m-lx} shows the rest-frame $V$-band and $K$-band
luminosities from our SED fits against the X-ray luminosities from
\citet{alexanderb}.  There is an apparent trend between $K$-band 
and X-ray
luminosity, which is reasonably fit by a power-law of the form
L$_K\propto$\,L$_X^{0.64\pm0.06}$.  To test the strength of the
correlation, we applied a Spearman rank test, The correlation 
coefficient is 0.68, with a 2.4$\sigma$ deviation from the null-hypothesis
that L$_K$ and L$_X$ are not correlated.  Removing the three lower
redshift sources in 
Fig.~\ref{fig:m-lx} does weaken the apparent correlation: 
the power law index becomes $0.79\pm0.32$, while
the Spearman test estimates a 1.8$\sigma$ deviation from the
null-hypothesis.  While the
modest dynamic range of our sample makes evidence for a 
correlation difficult to find, as we discuss below, 
it does yield strong
constraints on the normalization of the relationship.

For both the full sample and with the three
lower redshift sources excluded, the trends 
with $K$-band magnitude are three times tighter
than that seen between the reddening-corrected rest-frame $V$-band 
magnitude and
X-ray luminosities, yielding an RMS of 0.6 versus 2.0\,mags
respectively. This difference most likely reflects the increased
sensitivity to reddening corrections in the rest-frame $V$-band, in 
addition to the fact that near-infrared luminosities better probe the
evolved stellar population.
Since the uncertainty in the stellar mass estimates scale
as $0.4\Delta_{\rm mag}$ dex, these results underline the advantages
of the rest-frame $K$-band luminosities to estimate the stellar masses
of these galaxies.  

If we assume the X-ray emission arises from
accretion onto SMBHs, as indicated by the spectral analysis of
\citet{alexanderb}, we can interpret the correlation between $L_K$ and
$L_X$ as one between the stellar component of the galaxies and their
SMBHs.  Such a correlation would also be expected 
if the $K$-band light were due to hot dust powered by the AGN.  However,
with only a few exceptions, the SEDs seem to be dominated by stellar light, 
as evident from a well-defined stellar bump peaking near 1.6\mum  
(Fig.~\ref{fig:allseds}).  
For the remainder of the paper we will assume that the $K$-band is 
tracing the stellar mass, but the reader should bear the caveat in mind.

\begin{figure}
\plotone{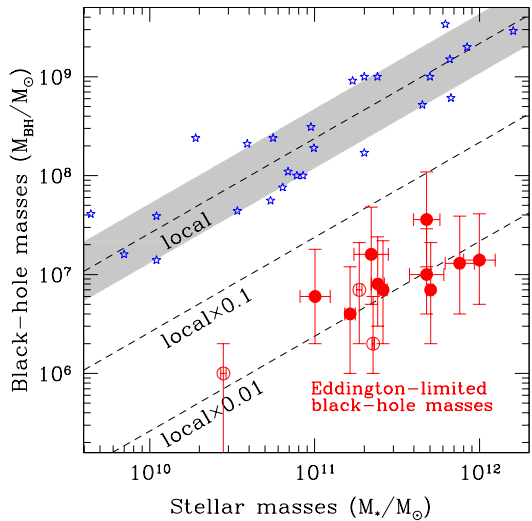}
\caption{The relation between stellar and SMBH masses for our sample of 13 
high-redshift SMGs.  Black hole masses are derived using the assumption of 
Eddington-limited accretion (see \S\ref{sec:bhmass}).
Stellar masses are derived by using the rest-frame $K$-band absolute 
magnitude derived from the fits to the SEDs in Fig.~\ref{fig:allseds} and 
assuming L$_K$/M=3.2 for all galaxies.
The upper dashed line denotes the best fit to local galaxies (shown as stars)
as derived by \citet{2003ApJ...589L..21M}, and the grey area denotes the 
region where 68\% of the local points are contained.  The lower dashed
lines denote tracks with black hole masses a tenth and hundredth of the 
local relation.  The three SMGs shown as open symbols are those
lying at redshifts of $z<1.8$.
}
\label{fig:magorrian}
\end{figure}

In Fig.~\ref{fig:magorrian} we plot the distribution of
M$_{\star}$ -- the
stellar mass from the SED fits assuming a fixed L$_K$/M,
and M$_{\rm BH,Edd}$ -- the SMBH masses derived from the absorption-corrected
X-ray luminosities by \citet{alexandera} assuming Eddington accretion
and a $6^{12}_{-4}$\% bolometric correction. As in Fig.~\ref{fig:m-lx},
we see a reasonable correlation between these two quantities. 

The limited dynamic range of our sample and the uncertainties in the
individual measurements means we cannot provide a reliable constraint
on the power-law index for this correlation, M$_{\rm
  BH}\propto$\,M$_{\star}^\gamma$. The data are consistent with
$\gamma\sim $0.3--1 with a very large uncertainty depending upon the
exact choice of sample used.  Nevertheless, while the slope of the
correlation is uncertain, the normalization is well-constrained at the
mean M$_{\star}$ of our sample. At $\log({\rm M_{\star}})=11.4\pm0.4$, we
find that the average Eddington-limited black-hole mass is
$\log($M$_{\rm BH,Edd})=6.86\pm0.40$.  If we exclude the three SMGs at
$z<1.8$, the mean stellar and black hole masses are slightly larger:
$\log({\rm M_{\star}})=11.5\pm0.32$ and $\log($M$_{\rm
  BH,Edd})=7.00\pm0.27$

As a comparison we turn to a sample of nearby
\mbox{($<250$\,Mpc)} galaxies with reliable black-hole masses estimated from
high resolution \hst\ spectroscopy, and stellar masses derived via
near-infrared observations of the host galaxies
\citep{2003ApJ...589L..21M}.  Although their full sample includes 37 galaxies,
we concentrate on the 27 ``Group 1'' systems judged by those authors to be the
most secure.  Using these data, \citet{2003ApJ...589L..21M} showed that:

\begin{eqnarray}
\lefteqn{\log({\rm M_{BH}}) = (8.28\pm0.06) + } & &\\
&\mbox{\hspace{1.0in}} (0.96\pm0.07)\log({\rm M_{Bulge}}/10^{10.9}).&\nonumber
\label{equ:mag}
\end{eqnarray}
We note that this relation predicts black hole masses roughly 50\% higher
than those presented in
\citet{2004ApJ...604L..89H},
who used Jeans equation modeling rather than virial 
estimators to predict the bulge mass.  

Using these relations, our data suggests 
that the black hole masses we infer for SMGs  (either in the full
sample or for the ten SMGs at $z>1.8$)
are $\sim50$ times smaller than for similarly massive local spheroids.

\section{Discussion}
With the assumptions about the L$_K$/M and Eddington
fraction $\eta$ we have adopted, 
our results suggest that the black holes in SMGs 
require significant growth in order to approach the present-day
relation.  To understand the significance of this conclusion, however, we
need to better characterize any biases in the
mass estimates for both  of these components. Note that errors on
individual sources are large in principle, since for both the black-hole
and stellar masses we have assumed ensemble values (i.e.\ fixed $\eta$
for the black-holes, and a single L$_K$/M value for all stellar mass
estimates).  It is quite likely that source-to-source variations
will affect the scatter of the data, but until more accurate methods
are available for the mass estimates, we concentrate instead on 
understanding any systematic bias which could effect the sample as a whole.

\begin{figure}
\plotone{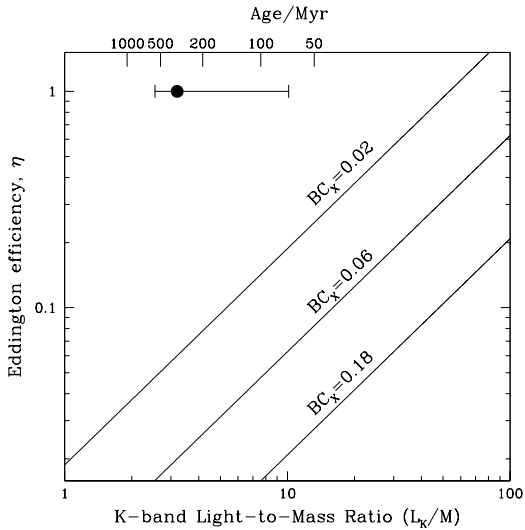}
\caption{This figure graphically illustrates the available parameter
space required to place the SMBH and
stellar masses of the high-redshift
SMG sample onto the local M$_{\ast}$-M$_{\rm BH}$ relation.
The solid lines represent combinations of L$_K$/M and $\eta$
that would bring our mass estimates in line with the local relation
at a given assumed value for the X-ray bolometric correction. 
They span
the estimate for BC$_X$ determined by \citet{1994ApJS...95....1E}. 
We plot our
best estimate for the values of L$_K$/M and $\eta$ as the solid point. 
As an
example to help illustrate
the use of this figure: 
assume a 6\% X-ray to bolometric luminosity correction
and an Eddington efficiency of 0.1, the plot indicates that our stellar ages
need to be $\sim50$\,Myr in order to make the data agree with the local 
relationship.  The plot does not account for a possible missing fraction
of X-ray flux due to heavy obscuration in the SMGs.  This would systematically
push the black-hole mass estimates up, and the three lines closer to our
observed point.  See \S\ref{sec:bhmass} and \citet{alexanderb} for more
details.
}
\label{fig:range}
\end{figure}

Altogether, the ratio between between our black hole masses and those
in the nearby Universe scales as:
\begin{equation}
\label{equ:unity}
50\left({\eta}\over{1.0}\right)\left({3.2}\over{{\rm L}_K/{\rm M}}\right)\left({{\rm BC}_X}\over{0.06}\right),
\end{equation}
where we
have simplified the expression by assuming the power-law relation between
stellar and black hole has an exponent of 1.0.  
In Fig.~\ref{fig:range}
we plot tracks of different conditions where the above expression is unity.
This demonstrates how difficult it is for
any one factor to explain the difference between our
sample and the local relation, and suggests that either the 
discrepancy is real, 
or that biases in the parameters must conspire if, in fact, SMGs have a 
black-hole/stellar mass ratio comparable to local systems.  We now
discuss these potential biases in more detail.

\subsection{Factors influencing stellar mass estimates}
We begin by pointing out that  the stellar masses we
derived are
on average five times higher than those reported from
modeling the UV/optical SEDs of 
SMGs in \citet{2004ApJ...616...71S}. This contrasts with the 
reasonable agreement in
the estimated stellar masses 
derived for UV-selected galaxies at similar redshifts
to the SMGs, by Shapley et al.\ (2005) from both
rest-frame UV/optical and rest-frame UV/optical/near-infrared 
photometry.   We attribute the difference 
in our masses to those from  \citet{2004ApJ...616...71S}
to the absence of mid-infrared 
information to constrain the SED
modeling in the earlier work, which is then much more sensitive
to the reddening corrections derived from the SED fits 
to these very dusty galaxies and the
possibilities of strong differential reddening within their
young stellar
populations.  

In our modeling of the stellar mass of the systems we adopted
a L$_K$/M ratio of 3.2, characteristic of a burst with
an age of $\sim 250$\,Myrs.  If we require that the SMG and the
present-day M$_{\star}$--M$_{\rm BH}$ relations agree, then 
from Eq.~\ref{equ:unity} we see that the stellar masses 
must drop by a factor of $\sim50$.
This corresponds to L$_K$/M$\sim250$, which is not possible
with the Miller-Scalo IMF.  For a Salpeter IMF, such a high 
L$_K$/M ratio implies ages for the SMGs of $\lsim 10$\,Myrs. 
We discuss the implications for varying the IMF later in this
section.

However, we note that there are fundamental inconsistencies in our
calculations if we rigidly interpret the star formation histories we used.
The SMGs in our sample have been selected on the basis of their intense
far-infrared luminosities, which implies high star formation rates. 
The fast decline in the far-infrared emission
from a starburst after star formation ceases \citep{1998ARA&A..36..189K} 
indicates
that star formation is currently on-going in these systems or has declined
significantly in only the last few 10's of Myrs.  The submm detection of 
these sources is therefore formally inconsistent with the star formation
history in our burst model -- where there has been no star formation
within the galaxy in the past $\sim250$\,Myrs.  Although the
alternative CSF models include on-going star formation, the much higher
ages derived from these fits are equally inconsistent with a constant rate
of star formation at the level indicated by the observed far-infrared
emission from the SMGs, $\sim$\,1--2\,$\times 10^3$\,M$_\odot$\,yr$^{-1}$
\citep{1998ARA&A..36..189K}.  
To form an $M_K\sim -26.4$ galaxy with a constant SFR
at this level would take just $\sim$\,100\,Myrs, not the 2\,Gyrs derived
from the fits.

There are two solutions to solve these inconsistencies within our models:
1) allow a more complex star formation history, most likely involving a
recent burst of activity; 2) apply age-dependent dust extinction to the
stellar populations, with the youngest stars being most extincted. As both
solutions involve additional free-parameters to what is already a
weakly-constrained problem, we will only qualitatively discuss their
impact.

Bursts of star formation are a natural consequence of interaction-induced
activity due to the cyclic behavior of bound orbits 
\citep{1994ApJ...431L...9M}.
They lead to a dispersion in the properties of SMGs dependent on
the number of previous bursts they have experienced.  The main consequence
in terms of our analysis is the realization that the present activity may
not have produced the total stellar population we observe.  Mixing in a
fraction of older stars will systematically decrease the effective
L$_K$/M, potentially increasing the masses of the galaxies.  This would
therefore push the SMGs even further away from the $z=0$
M$_{\star}$--M$_{\rm BH}$ relation.

Age-dependent or selective extinction of the stellar populations 
\citep[e.g.][]{2001ApJ...550..195P} within SMGs can preferentially
extinguish the UV emission from the youngest stars.  This results in SEDs
with much redder spectral indices between the UV and near-infrared than
can be achieved with a simple foreground dust screen acting upon the whole
stellar population.  When the diffusion timescale for stars to escape
their obscured birth-places is around a few 100's Myrs, this process can
also help reproduce the strength of the spectral breaks at
$\sim$\,4000\AA, even in galaxies with significant on-going star
formation.  As a result, selective extinction would provide the
opportunity to fit the SED of our SMG sample with relatively young CSF
models, $\sim$\,100\,Myrs, consistent with the ages needed to produce
their observed near-infrared luminosities given the star formation rates
inferred from their far-infrared emission.

Attempts to estimate 
typical ages for SMGs include the spectral dating
of the few examples with high-quality UV spectroscopy.
Such an analysis by \citet{2003MNRAS.342.1185S} indicates an age for a
UV-bright SMG of $\gsim 10$\,Myrs, similar to the minimum age we
require.  However, as \citet{2003MNRAS.342.1185S} state the SMG 
they model may be 
a rare system and the small proportion of SMGs showing similar signs of
young stellar populations in their spectra suggests a more
typical age for an SMG would be 50--100\,Myrs.  
Another estimate of the likely age of SMGs can be derived via
gas-depletion timescales using the measured mass of the gas reservoir
and the estimated star formation rates suggest likely lifetimes of
$\gsim 30$\,Myrs \citep{gravy}, where we have
assumed an AGN contribution to
the far-infrared luminosities in line with \citep{alexanderb}.
Finally, 
taking our estimated stellar masses and the median star formation rates
derived from the far-infrared luminosities of the SMGs in our sample using our
adopted IMF, we determine lifetimes of $\sim 100$\,Myrs to produce the
stellar populations we see.   Balancing these estimates with those  from
the SED fitting, we conclude that the typical ages of
SMGs are likely to be $\gsim 50$\,Myrs.  Hence 
the stellar masses are inconsistent with what would be predicted
from the local M$_{\star}$--M$_{\rm BH}$ relation assuming our black-hole
masses are accurate.

Are there other ways to change the effective L$_K$/M  without invoking
such young ages?  One way to reduce it is to have a 
significant contribution to the
rest-frame $K$-band light from the AGN.  
However the bulk of our sample, as evident in 
Fig.~\ref{fig:allseds}, are well-fit by a stellar component only.  
In the few cases where AGN cause the SED
fits to diverge at the longer wavelengths, the estimated luminosity
we derive at 2.2$\mu$m is still dominated by the stars.   We used
the 8\mum\ excess (the ratio of the measured 8\mum\ flux over that
derived from the best fit model template) to quantify possible AGN 
contamination.  We find that the $z>1.8$ sources (excluding the 3 
marked as having an obvious near-IR excess) have an average excess of
1.0, suggesting little or no contribution from an AGN.  The remaining three
sources however, have an average excess of $2.3$, which means a nearly equal 
contribution to the rest-frame $K$-band from AGN as stars. Scaling down the
stellar masses of these three galaxies to account for this results in
 only a $\sim0.1$ reduction in the mean stellar mass of the sample, which
is small considering the other factors we discuss.

In addition to contribution of light from the AGN, we note that our 
analysis includes light from {\it all} of the stars, and not
simply the bulge component used in deriving the local relation.  While the 
high-resolution \hst/NICMOS imaging verifies
that the rest-frame $V$-band light in SMGs 
are reasonably fit by a $r^{1/4}$ surface 
brightness profile \citep{boryshighres}, some systems do show a 
significant disk
component. Whether these stars are still in a disk component  at the
present-day or whether they would have been driven into the bulge
through, for example,
bar-instabilities, is impossible to verify, and so we flag this to
the reader as a fundamental uncertainty of our analysis.  Nevertheless,
we note that any correction to the stellar masses due to a disk
component is unlikely to change the inferred spheroid mass by more than
a factor of 2--$3\times$ for a small fraction of the SMGs.

The IMF also has a significant effect on stellar mass estimates.
\citet{2005MNRAS.356.1191B} argue that an IMF biased to more
massive stars is required to reproduce the properties of SMGs in
current semi-analytic galaxy formation models, and
invoke a power-law IMF with an index flatter than the 
canonical Salpeter 
\citep{1955ApJ...121..161S} slope of $\alpha\sim2.35$.   We used 
{\sc Starburst99} to study the effect of this top-heavy IMF on our
mass estimates, but find that the stellar mass estimates change by at most 
a factor of $\sim3$ for the range of likely ages for SMGs.  An 
additional factor of three in the
light-to-mass ratio can be picked up by truncating the IMF at 
$\sim1\,$M$_\odot$, and there are different IMF prescriptions (such as the
one proposed by \citet{maraston} that has an increased $K$-band
contribution from thermally pulsating AGB stars) that 
can also increase the effective light-to-mass ratio.  Altogether
it is possible that these effects could conspire to decrease our
stellar masses by up to a factor of $\sim10$.  However,
we conclude that any bias present in the stellar
mass is insufficient by itself to account for the difference 
between our results and the local relation.

\subsubsection{Tests of the stellar masses}
Are the high stellar masses we derive in conflict with other estimates
of the masses of SMGs?

Dynamical masses for SMGs from CO spectroscopy with millimeter
interferometers
indicate a median gas+stellar masses of 
$\log({\rm M_{dyn}})\sim 11.1\pm 0.3$
within the central $\lsim 10$\,kpc of
typical SMGs with 850-$\mu$m fluxes of a few mJys
\citep{1998ApJ...506L...7F,1999ApJ...514L..13F,2003ApJ...597L.113N,gravy}. 
In addition,
five of the galaxies in our sample have dynamical mass estimates from
H$\alpha$ spectroscopy \citep{2004ApJ...617...64S}, which all
give constraints on the masses of $\log({\rm M_{dyn}})\sim11.0$--11.3,
again on $\sim 10$\,kpc scales.  
In a recent analysis, \citet{gravy} use the integrated CO line
luminosities
for  12 SMGs to estimate an average contribution from gas of
$\log({\rm M_{gas}})=10.5\pm0.2$ within $\lsim 10$\,kpc
for these systems.  Thus taking into account the contribution
from gas, the dynamical estimates are 
consistent with the presence of $\log(M_*)\sim 11$ of stars within
the central regions of the SMGs.

As highlighted by \citet{2004ApJ...617...64S} the H$\alpha$ dynamical
mass estimates derived from the H$\alpha$ line
widths for SMGs
are five times higher than similar constraints on UV-selected galaxies
at comparable redshifts \citep{2003ApJ...591..101E}.  To see if
this difference is also reflected in the stellar mass
estimates for the two populations, we ran our \hz\ analysis on the sample of 
72 spectroscopically confirmed, UV-selected, and IRAC-detected, galaxies at 
$z\sim 2.3$ presented in \citet{aliceirac}.  These UV-selected 
galaxies 
\citep{2004ApJ...607..226A,2003ApJ...592..728S} 
make an ideal comparison set, as they are more modestly star-forming galaxies
at roughly the median redshift of our sample.
As with our analysis, the fits to the
data from \citet{aliceirac} were degenerate in SED template, 
age and reddening.  
Roughly one-third of the objects are best fit by a young ($\sim50\,$Myr) burst 
while the remainder are better modeled by a constant star-formation 
history with age of $\sim600$\,Myr.  Again we find that the choice of 
model has little impact on the
derivation of the stellar mass, and we obtain a mean of 
$\log(\rm{M_*})=10.35\pm0.34$
for this sample.  The estimated ages and masses are in 
good agreement to the results which \citet{aliceirac} derive
using a more detailed analysis ($\log(\rm{M_*})=10.32\pm0.51$).  
This confirms that our approach to stellar mass estimation is robust 
and also suggests that the SMGs in our sample are on average $\sim10$ 
times more massive than the UV-selected population at $z\sim 2.2$.

Overall, we conclude that our stellar masses are consistent with 
dynamical limits on the masses of the SMG population and support the
suggestion that the 
central regions of these galaxies are baryon dominated.  We note
that the CO detections of SMGs indicate
that there is 
sufficient gas present to continue to build
the black hole, but at the same time the gas masses are small enough
that they will not appreciably change the stellar masses even if the bulk of
it is converted into stars.  Only by bringing in more gas
or stars into these
systems (through cooling or
mergers) can their stellar masses be significantly increased from what
we measure.

\subsection{Factors influencing black hole mass estimates}
\label{sec:bhmass}
Is it possible that the sources only appear to lie 
below the local relation because
the black-hole mass estimates from \citet{alexandera} are too low? 
Certainly, in the light of the discussion above it appears that
the black hole mass estimates are probably the more important source
of uncertainty in our analysis.  In
estimating the black-hole masses, \citet{alexandera} assumed
Eddington-limited accretion and a $6^{+12}_{-4}$\% bolometric
correction to convert the observed absorption-corrected L$_X$ to
M$_{\rm BH,Edd}$. Since the Eddington limit appears to provide a
reasonable upper bound to the accretion rate of black holes 
\citep[e.g.,][]{2004MNRAS.352.1390M}, this approach essentially provides a
lower limit on the black-hole masses and sub-Eddington accretion would
imply larger black holes. However, an average black-hole mass $\sim50$
times higher is needed to reconcile our data at the characteristic
stellar mass of the SMGs, $\sim 10^{11}$M$_\odot$.  This would require
$\eta\sim0.02$, which is low given the gas-rich environments in
which the SMBHs are likely to reside. Although we do not know the
Eddington rate of these galaxies, the large fraction of SMGs that host
AGN activity suggests that the accretion is likely to be reasonably
efficient (see \citet{alexandera}). There is tangential support for
this claim from studies that have estimated black-hole masses for
other active high-redshift populations independently of the Eddington rate,
finding $\eta\sim$\,0.1--1.7 \citep[e.g.][]{marconi04,
2004MNRAS.352.1390M,barger05}. At the smallest
accretion efficiencies found in these studies, the black-hole masses
would still be $\sim$\,5 times too small for our estimated stellar
masses.

There is further support for Eddington-limited or near Eddington-limited
accretion from the similarity between the optical/near-infrared spectra of
some SMGs with those of local Narrow Line Seyfert 1 (NLS1) galaxies
\citep{1998MNRAS.298..583I,2001A&A...380..409V,
  2002ApJ...577L..79L,2003MNRAS.342.1185S,2003AJ....125.1236D,
  2004ApJ...617...64S,2005MNRAS.359..401S}; see \S4.3 of
\citet{alexanderb}. If the spectral similarities of the AGN signatures
in these two populations are suggestive of comparable physical
conditions in the accretion disk around the SMBH, then the
approximately Eddington-limited rates estimated for local NLS1s may be
appropriate for SMGs
\citep[e.g.][]{2002ApJ...565...78B,2004A&A...426..797C}. Moreover,
NLS1s appear to have smaller black holes than those predicted from the
local M$_{\star}$--M$_{\rm BH}$ relation; Using  reverberation
mapping, \citet{2004ApJ...606L..41G}
and \citet{2004AJ....127.3168B} show that
that the SMBHs in NLS1s are an order of magnitude smaller
than predicted for galaxies having a stellar mass of
$10^{11}$M$_\odot$.  Thus physical models for the accretion disks and
SMBHs in NLS1s may be very relevant for understanding the growth of
SMBHs in massive, young galaxies at high redshifts.  We do caution the
reader that the broad-line region in SMGs
may be missed due to heavy extinction in
these dusty galaxies, and hence the spectra may be dominated by a less
obscured narrow-line region.  High signal-to-noise spectra would be useful
to search for faint, broad wings that may be otherwise  missed.

There is also theoretical support for $\eta\sim1$ efficiencies:
\citet{2004ApJ...600..580G} and \citet{2003ApJ...583...85K} predict
that this condition will be met near the peak epoch of star formation,
where these objects have a stronger chance of being detected at submm
wavelengths.  Interestingly, these models also predict black hole
masses $\approx$~1--2 orders of magnitude smaller than the local
relation during the obscured, star-bursting phase, similar to what our
data suggest.  Hence, while the assumption of Eddington limited accretion 
is a large one, it is not a unreasonable choice.  We reiterate that
even an efficiency 10 times smaller would still result in the
black-holes in SMGs being smaller
than would be predicted by the local relation.

Another uncertainty in the SMBH estimates comes from the bolometric
correction applied to the X-ray luminosities. \citet{alexandera}
assumed that $6^{+12}_{-4}$\% of the energy output due to accretion
was emitted in the X-ray waveband \citep{1994ApJS...95....1E}, but if
this was lower, the inferred mass of the SMBH would rise.  However,
they also pointed out that this correction is similar to the estimated
bolometric corrections for obscured and unobscured AGN of similar
luminosity to the AGNs in the SMGs \citep[see][]{alexandera}.

A final potential source of uncertainty in the SMBH masses comes from
the possibility that a substantial fraction of the X-ray emission from
the SMBH is still obscured at rest-frame energies of $\sim$\,20\,keV.
For instance, \citet{2005MNRAS.357.1281W}
find that the \chandra\ 2-Ms exposure of the CDF-N fails to recover 
approximately 40\% of the hard X-ray background. This may in part be
hidden within highly-obscured `Type-2' quasars \citep{type2}, 
but dusty SMGs also harbor obscured AGN (and could in some
cases be Type-2 quasars).  Indeed, not all SMGs are detected in
the \chandra\ 2-Ms X-ray image and so potentially some fraction of
the X-ray emission from the SMBH in this population could be missed
from the \chandra\ analysis.  In \S{3.5} of \citet{alexanderb}, it
is pointed out that while  obtaining tight constraints on the
obscuring column densities for individual sources is difficult, the
agreement between their models and X-ray
spectra of SMGs ranked by obscuration class is
excellent, suggesting that most of the X-ray emission in these sources
is accounted for.  If a large fraction of the X-ray emission in SMGs 
is still absorbed, we will need to wait for the next generation 
of X-ray observatories to find it.

\subsection{Implications for the evolutionary history of SMGs}
To summarize the previous two subsections,  it is possible to
find combinations of the parameters used to estimate
the stellar and black hole masses of SMGs which would
reconcile them with the local M$_{\star}$--M$_{\rm BH}$ relation.
For example, if  we 
assume a 6\% bolometric correction, then we require a combination
of a typical SMG age of just $\sim 50$\,Myrs and an Eddington efficiency of 
$\eta\sim 0.1$ to place the high-redshift data on the local relation
(Fig.~\ref{fig:range}). 
This combination is plausible, and if it is the case then the implication
is that the stellar and black hole mass grow simultaneously, with a mass 
ratio that is similar at high redshift as well as low.  However,
the need for both low Eddington accretion and a very short lifetime for the 
SMGs is uncomfortable and hence we should also explore the possibility 
that the black holes in SMGs are, in fact
less massive than expected from their stellar
masses and the low
redshift M$_{\star}$--M$_{\rm BH}$
relation. The following discussion assumes that this is indeed the case.

The theoretical models of \citet{2004ApJ...600..580G}
suggest that SMBHs may grow exponentially, taking only a few 10's of Myr 
to mature \citep{2003ApJ...583...85K,2004ApJ...600..580G}.  
This is a short enough time-scale that SMGs with
more evolved SMBHs, which lie on the
local relationship, should fall within the redshift range probed by 
our sample.  That we do not see such systems implies that as the black hole 
rapidly builds up its mass, it either shuts down the star-formation that
powers the far-infrared luminosities (hence they drop out of the
submillimeter samples)
or they become so obscured that they are no longer detected in the
X-ray waveband.

The former argument supports a scenario that links the QSO and SMG
populations.  As the black hole undergoes a rapid buildup of mass, the
ensuing strong radiation field strips away the obscuring gas and dust,
terminating star formation and eventually 
allowing the AGN to be visible for a short time as a luminous optical quasar
\citep{1988ApJ...325...74S,1998A&A...331L...1S,
1999MNRAS.308L..39F,2005Natur.433..604D,alexanderb}.
The close similarity between
the redshift distribution of radio-identified SMGs and QSOs 
\citep{chapmanrsmg} also lends  support to this hypothesis,
as does the relative space densities of SMGs and QSOs given
our estimates of their respective lifetimes
\citep{2004ApJ...616...71S}.
In addition,
recent high resolution adaptive optics
imaging by 
\citet{qsoao} has shown that 
the black hole masses and host galaxy luminosities of $z\sim2$
quasars are consistent with the local M$_*$--M$_{\rm BH}$ relation,
suggesting that  QSOs could be the end-product of the rapid growth of SMBHs
seen soon after an SMG-like phase.  
Finally, 
work by \citet{2001Sci...294.2516P,2004ApJ...611L..85P} using SCUBA to 
measure the rest-frame far-infrared luminosities of
unobscured and moderately obscured QSOs at $z\sim 2$
verify that the former class 
represent a phase of significant SMBH growth but not bulge growth,
while the latter may be a transition phase between the
star-formation dominated SMGs and AGN-dominated quasars.
Thus, if SMGs are the progenitors of quasars, 
they {\it must} lie below the locally determined M$_{\star}$--M$_{\rm BH}$
relation.

In conclusion, there is good qualitative and quantitative agreement
for an evolutionary sequence which connects moderate
mass SMBHs in far-infrared luminous, star-forming galaxies at $z\sim 2$--3
(when most of the spheroid is produced), with subsequent 
growth of the SMBHs during a short phase as a luminous quasar to 
result in  massive spheroids at the present-day which lie on
the M$_{\star}$--M$_{\rm BH}$ relation.

\section{Conclusion}
Using deep \chandra, \spitzer, and ground-based optical
and near-infrared imaging data, we have 
estimated the stellar and black hole masses for a sample of 13 SMGs with
precise spectroscopic redshifts.  Our main conclusions are:
\begin{enumerate}

\item{There is a tentative
correlation between the rest-frame near-infrared and X-ray
luminosity in this sample of SMGs.  The X-ray
emission 
likely arises from accretion of matter onto a SMBH, and so can be used to estimate its mass.
We find a similar correlation between rest-frame X-ray and  $V$-band
luminosities, but this exhibits
more scatter, suggesting that it is more
sensitive to dust and age variations in these young, obscured
galaxies.}

\item{The restframe UV--optical--near-infrared
SEDs of most of
our SMGs are dominated by stellar emission and not light from the
  AGN, and are well-described by either a young  burst
model, or by an older galaxy with a more continuous star-formation
history.  In both cases, the derived
light-to-mass ratios, and hence stellar mass estimates, are similar.} 

\item{The mean stellar mass for the sample is 
$\log({\rm M_{\star}})=11.4\pm0.4$, consistent with
H$\alpha$ and CO dynamical mass estimates for the SMG
population.  Hence the \spitzer\
observations have shown that these intense star-forming galaxies
have a significant stellar population in place at $z\sim 2.2$.}

\item{We find that SMGs are 
typically 10 times more massive than typical UV-selected galaxies at
similar redshifts.}

\item{We use the assumption of Eddington-limited accretion to estimate the
black hole masses for each galaxy in the sample.  We find that for a given
stellar mass, the black holes are $\sim50$ times less massive than those
for spheroids with similar stellar masses in the local Universe.}

\item{In order to determine if the high-redshift SMGs lie of the local 
M$_{\star}$--M$_{\rm BH}$ relation, we explore the range of assumptions that are used to estimate 
both the stellar and black hole masses.  
We find that most reasonable combinations of parameters lead to
lower black hole masses at a fixed stellar mass in SMGs at $z\sim2$ compared
to the low redshift relation for spheroids.}

\item{We suggest the NLS1 galaxies may be a good template for understanding
the growth of SMBHs at high redshift.  They have spectral properties common 
with many SMGs, and (locally) have black hole masses smaller than predicted
for less active spheroids.}

\item{Our results are consistent with a model where the SMBHs in 
SMGs subsequently undergo rapid growth
to reach the local M$_{\star}$--M$_{\rm BH}$
relation.  Such a phase would most naturally be explained if the
SMGs  evolved into quasars on their way to becoming luminous, passive
ellipticals at the present-day.}

\end{enumerate}

\acknowledgments
We thank Christine Done, Duncan Farrah and
Rowena Malbon for useful conversations, and Mark Brodwin
for advice on IRAC photometry measurements.  We also thank
an anonymous referee for a constructive report which helped to
improve the structure and content of this paper.
IRS and DMA acknowledge 
support from the Royal Society. AWB is supported by NSF AST-0208527, 
the Sloan Foundation and the Research Corporation.


\begin{thebibliography}{}
\bibitem[Adelberger et al.(2004)]{2004ApJ...607..226A} Adelberger, K.~L., Steidel, C.~C., Shapley, A.~E., Hunt, M.~P., Erb, D.~K., Reddy, N.~A., \& Pettini, M.\ 2004, \apj, 607, 226 
\bibitem[Alexander et al.(2003)]{2003AJ....125..383A} Alexander, D.~M., et al.\ 2003, \aj, 125, 383
\bibitem[Alexander et al.(2005a)]{alexandera} Alexander, D.~M., Smail, I., Bauer, F.~E., Chapman, S.~C., Blain, A.~W.,  Brandt, W.~N., Ivison, R.~J. \ 2005a, \nat, accepted
\bibitem[Alexander et al.(2005b)]{alexanderb} Alexander, D.~M., Smail, I., Bauer, F.~E., Chapman, S.~C., Blain, A.~W.,  Brandt, W.~N., Ivison, R.~J. \ 2005b, \apj, accepted
\bibitem[Barger et al.(2000)]{barger05} Barger, A.~J., Cowie, L.~L., Mushotzky, R.~F., Yang, Y., Wang, W.-H., Steffen, A.~T., \& Capak, P.\ 2005, \aj, 129, 578
\bibitem[Barger et al.(2000)]{2000AJ....119.2092B} Barger, A.~J., Cowie, L.~L., \& Richards, E.~A.\ 2000, \aj, 119, 2092
\bibitem[Baugh et al.(2005)]{2005MNRAS.356.1191B} Baugh, C.~M., Lacey, C.~G., Frenk, C.~S., Granato, G.~L., Silva, L., Bressan, A., Benson, A.~J., \& Cole, S.\ 2005, \mnras, 356, 1191 
\bibitem[Benson et al.(2003)]{2003ApJ...599...38B} Benson, A.~J., Bower, R.~G., Frenk, C.~S., Lacey, C.~G., Baugh, C.~M., \& Cole, S.\ 2003, \apj, 599, 38 
\bibitem[Bertin \& Arnouts(1996)]{1996A&AS..117..393B} Bertin, E., \& Arnouts, S.\ 1996, \aaps, 117, 393 
\bibitem[Blain et al.(2004)]{2004ApJ...611..725B} Blain, A.~W., Chapman, S.~C., Smail, I., \& Ivison, R.\ 2004, \apj, 611, 725
\bibitem[Bolzonella et al.(2000)]{2000A&A...363..476B} Bolzonella, M., Miralles, J.-M., \& Pell{\' o}, R.\ 2000, \aap, 363, 476 
\bibitem[Boroson(2002)]{2002ApJ...565...78B} Boroson, T.~A.\ 2002, \apj, 565, 78 
\bibitem[Borys et al.(2004)]{2004MNRAS.355..485B} Borys, C., Scott, D., Chapman, S., Halpern, M., Nandra, K., \& Pope, A.\ 2004, \mnras, 355, 485
\bibitem[Borys et al.(2006)]{boryshighres} Borys, C., Chapman, S., Smail, I.~R., \& Ivison, R.~J.\ 2006, \apj, in prep.
\bibitem[Botte et al.(2004)]{2004AJ....127.3168B} Botte, V., Ciroi, S., Rafanelli, P., \& Di Mille, F.\ 2004, \aj, 127, 3168
\bibitem[Bundy et al.(2005)]{bundy} Bundy, K., Ellis, R.~S., \& Conselice, C.~J.\ 2005, \apj, accepted
\bibitem[Calzetti et al.(2000)]{2000ApJ...533..682C} Calzetti, D., Armus, L., Bohlin, R.~C., Kinney, A.~L., Koornneef, J., \& Storchi-Bergmann, T.\ 2000, \apj, 533, 682 
\bibitem[Capak et al.(2004)]{2004AJ....127..180C} Capak, P., et al.\ 2004, \aj, 127, 180
\bibitem[Cavaliere \& Vittorini(2002)]{2002ApJ...570..114C} Cavaliere, A., \& Vittorini, V.\ 2002, \apj, 570, 114
\bibitem[Chapman et al.(2005)]{chapmanrsmg} Chapman, S.~C., Smail, I., Blain, A.~W., \& Ivison, R.~J.\ 2005, \apj, accepted 
\bibitem[Chapman et al.(2004)]{2004ApJ...611..732C} Chapman, S.~C., Smail, I., Windhorst, R., Muxlow, T., \& Ivison, R.~J.\ 2004, \apj, 611, 732
\bibitem[Chapman et al.(2003)]{2003Natur.422..695C} Chapman, S.~C., Blain, A.~W., Ivison, R.~J., \& Smail, I.~R.\ 2003, \nat, 422, 695 
\bibitem[Collin \& Kawaguchi(2004)]{2004A&A...426..797C} Collin, S., \& Kawaguchi, T.\ 2004, \aap, 426, 797
\bibitem[Cox(2000)]{2000asqu.book.....C} Cox, A.~N.\ 2000, Allen's astrophysical quantities, 4th ed.~Publisher: New York: AIP Press; Springer, 2000.~Edited by by Arthur N.~Cox.~ ISBN: 0387987460
\bibitem[Dawson et al.(2003)]{2003AJ....125.1236D} Dawson, S., McCrady, N., Stern, D., Eckart, M.~E., Spinrad, H., Liu, M.~C., \& Graham, J.~R.\ 2003, \aj, 125, 1236 
\bibitem[Dickinson et al.(2005)]{dickinson} Dickinson, M, et al.\ 2005, in preparation
\bibitem[Di Matteo et al.(2005)]{2005Natur.433..604D} Di Matteo, T., Springel, V., \& Hernquist, L.\ 2005, \nat, 433, 604 
\bibitem[Elvis et al.(1994)]{1994ApJS...95....1E} Elvis, M., et al.\ 1994, \apjs, 95, 1
\bibitem[Erb et al.(2003)]{2003ApJ...591..101E} Erb, D.~K., Shapley, A.~E., Steidel, C.~C., Pettini, M., Adelberger, K.~L., Hunt, M.~P., Moorwood, A.~F.~M., \& Cuby, J.\ 2003, \apj, 591, 101 
\bibitem[Fabian(1999)]{1999MNRAS.308L..39F} Fabian, A.~C.\ 1999, \mnras, 308, L39 
\bibitem[Farrah et al.(2005)]{farrah_stis} Farrah, D., Surace, J.~A., Veilleux, S., Sanders, D.~B., \& Vacca, W.~D. \ 2005,\apj, accepted
\bibitem[Frayer et al.(1999)]{1999ApJ...514L..13F} Frayer, D.~T., et al.\ 1999, \apjl, 514, L13
\bibitem[Frayer et al.(1998)]{1998ApJ...506L...7F} Frayer, D.~T., Ivison, R.~J., Scoville, N.~Z., Yun, M., Evans, A.~S., Smail, I., Blain, A.~W., \& Kneib, J.-P.\ 1998, \apjl, 506, L7
\bibitem[Gebhardt et al.(2000)]{2000ApJ...543L...5G} Gebhardt, K., et al.\ 2000, \apjl, 543, L5 
\bibitem[Giavalisco et al.(2004)]{2004ApJ...600L..93G} Giavalisco, M., et al.\ 2004, \apjl, 600, L93
\bibitem[Granato et al.(2004)]{2004ApJ...600..580G} Granato, G.~L., De Zotti, G., Silva, L., Bressan, A., \& Danese, L.\ 2004, \apj, 600, 580 
\bibitem[Greve et al.(2004)]{2004MNRAS.354..779G} Greve, T.~R., Ivison, R.~J., Bertoldi, F., Stevens, J.~A., Dunlop, J.~S., Lutz, D., \& Carilli, C.~L.\ 2004, \mnras, 354, 779
\bibitem[Greve et al.(2005)]{gravy} Greve, T.~R., et al.\ 2005, \mnras, accepted 
\bibitem[Grupe \& Mathur(2004)]{2004ApJ...606L..41G} Grupe, D., \& Mathur, S.\ 2004, \apjl, 606, L41
\bibitem[H{\" a}ring \& Rix(2004)]{2004ApJ...604L..89H} H{\" a}ring, N., \& Rix, H.\ 2004, \apjl, 604, L89 
\bibitem[Ivison et al.(1998)]{1998MNRAS.298..583I} Ivison, R.~J., Smail, I., Le Borgne, J.-F., Blain, A.~W., Kneib, J.-P., Bezecourt, J., Kerr, T.~H., \& Davies, J.~K.\ 1998, \mnras, 298, 583 
\bibitem[Ivison et al.(2002)]{2002MNRAS.337....1I} Ivison, R.~J., et al.\ 2002, \mnras, 337, 1 
\bibitem[John(1988)]{1988A&A...193..189J} John, T.~L.\ 1988, \aap, 193, 189 
\bibitem[Kaspi et al.(2000)]{2000ApJ...533..631K} Kaspi, S., Smith, P.~S., Netzer, H., Maoz, D., Jannuzi, B.~T., \& Giveon, U.\ 2000, \apj, 533, 631 
\bibitem[Kawakatu et al.(2003)]{2003ApJ...583...85K} Kawakatu, N., Umemura, M., \& Mori, M.\ 2003, \apj, 583, 85
\bibitem[Kennicutt(1998)]{1998ARA&A..36..189K} Kennicutt, R.~C.\ 1998, \araa, 36, 189
\bibitem[Kormendy \& Richstone(1995)]{1995ARA&A..33..581K} Kormendy, J., \& Richstone, D.\ 1995, \araa, 33, 581
\bibitem[Kuhlbrodt et al.(2005)]{qsoao} Kuhlbrodt, B., Orndahl, E., Wisotzki, L., \& Jahnke, K., \aap, submitted
\bibitem[Ledlow et al.(2002)]{2002ApJ...577L..79L} Ledlow, M.~J., Smail, I., Owen, F.~N., Keel, W.~C., Ivison, R.~J., \& Morrison, G.~E.\ 2002, \apjl, 577, L79 
 
\bibitem[Leitherer et al.(1999)]{1999ApJS..123....3L} Leitherer, C., et al.\ 1999, \apjs, 123, 3 
\bibitem[Magorrian et al.(1998)]{1998AJ....115.2285M} Magorrian, J., et al.\ 1998, \aj, 115, 2285
\bibitem[Maraston et al.(2005)]{maraston} Maraston, C. \ 2005, \mnras, submitted
\bibitem[Marconi et al. (2004)]{marconi04} Marconi, A., Risaliti, G., Gilli, R., Hunt, L.~K., Maiolino, R., \& Salvati, M.\ 2004, \mnras, 351, 169
\bibitem[Marconi \& Hunt(2003)]{2003ApJ...589L..21M} Marconi, A., \& Hunt, L.~K.\ 2003, \apjl, 589, L21 
\bibitem[Martinez-Sansigre et al.(2005)]{type2} Martinez-Sansigre, A., Rawlings, S., Lacy, M., Fadda, D., Marleau, F.~R., Simpson, C., Willott, C.~J., \& Jarvis, M.~J. \ 2005, \nat, in press 
\bibitem[McLure \& Dunlop(2004)]{2004MNRAS.352.1390M} McLure, R.~J., \& Dunlop, J.~S.\ 2004, \mnras, 352, 1390 
\bibitem[Mihos \& Hernquist(1994)]{1994ApJ...431L...9M} Mihos, J.~C., \& Hernquist, L.\ 1994, \apjl, 431, L9
\bibitem[Miller \& Scalo(1979)]{1979ApJS...41..513M} Miller, G.~E., \& Scalo, J.~M.\ 1979, \apjs, 41, 513 
\bibitem[Neri et al.(2003)]{2003ApJ...597L.113N} Neri, R., et al.\ 2003, \apjl, 597, L113
\bibitem[Page et al.(2001)]{2001Sci...294.2516P} Page, M.~J., Stevens, J.~A., Mittaz, J.~P.~D., \& Carrera, F.~J.\ 2001, Science, 294, 2516
\bibitem[Page et al.(2004)]{2004ApJ...611L..85P} Page, M.~J., Stevens, J.~A., Ivison, R.~J., \& Carrera, F.~J.\ 2004, \apjl, 611, L85 
\bibitem[Poggianti \& Wu(2000)]{2000ApJ...529..157P} Poggianti, B.~M., \& Wu, H.\ 2000, \apj, 529, 157 
\bibitem[Poggianti et al.(2001)]{2001ApJ...550..195P} Poggianti, B.~M., Bressan, A., \& Franceschini, A.\ 2001, \apj, 550, 195 
\bibitem[Salpeter(1955)]{1955ApJ...121..161S} Salpeter, E.~E.\ 1955, \apj, 121, 161 
\bibitem[Sanders et al.(1988)]{1988ApJ...325...74S} Sanders, D.~B., Soifer, B.~T., Elias, J.~H., Madore, B.~F., Matthews, K., Neugebauer, G., \& Scoville, N.~Z.\ 1988, \apj, 325, 74 
\bibitem[Shapley et al.(2005)]{aliceirac} Shapley, A.~E., Steidel, C.~C., Erb, D.~K., Reddy, N.~A., Adelberger, K.~L., Pettini, M., Barmby, P.,  \& Huang, J., \ 2005, \apj, accepted 
\bibitem[Shapley et al.(2001)]{2001ApJ...562...95S} Shapley, A.~E., Steidel, C.~C., Adelberger, K.~L., Dickinson, M., Giavalisco, M., \& Pettini, M.\ 2001, \apj, 562, 95 
\bibitem[Silk \& Rees(1998)]{1998A&A...331L...1S} Silk, J., \& Rees, M.~J.\ 1998, \aap, 331, L1 
\bibitem[Simpson \& Eisenhardt(1999)]{1999PASP..111..691S} Simpson, C., \& Eisenhardt, P.\ 1999, \pasp, 111, 691
\bibitem[Smail et al.(2004)]{2004ApJ...616...71S} Smail, I., Chapman, S.~C., Blain, A.~W., \& Ivison, R.~J.\ 2004, \apj, 616, 71 
\bibitem[Smail et al.(2003)]{2003MNRAS.342.1185S} Smail, I., Chapman,
  S.~C., Ivison, R.~J., Blain, A.~W., Takata, T., Heckman, T.~M.,
  Dunlop, J.~S., \& Sekiguchi, K.\ 2003, \mnras, 342, 1185 
\bibitem[Smail et al.(2002)]{2002MNRAS.331..495S} Smail, I., Ivison, R.~J., Blain, A.~W., \& Kneib, J.-P.\ 2002, \mnras, 331, 495
\bibitem[Smail et al.(1997)]{1997ApJ...490L...5S} Smail, I., Ivison, R.~J., \& Blain, A.~W.\ 1997, \apjl, 490, L5 
\bibitem[Steidel et al.(2003)]{2003ApJ...592..728S} Steidel, C.~C., Adelberger, K.~L., Shapley, A.~E., Pettini, M., Dickinson, M., \& Giavalisco, M.\ 2003, \apj, 592, 728
\bibitem[Steidel et al.(2004)]{steidel04} Steidel, C.~C., Shapley, A.~E., Pettini, M., Adelberger, K.~L., Erb, D.~K., Reddy, N.~A., \& Hunt, M.~P.\ 2004, \apj, 604, 534 
\bibitem[Swinbank et al.(2004)]{2004ApJ...617...64S} Swinbank, A.~M., Smail, I., Chapman, S.~C., Blain, A.~W., Ivison, R.~J., \& Keel, W.~C.\ 2004, \apj, 617, 64
\bibitem[Swinbank et al.(2005)]{2005MNRAS.359..401S} Swinbank, A.~M., et al.\ 2005, \mnras, 359, 401 
\bibitem[van Dokkum et al.(2003)]{2003ApJ...587L..83V} van Dokkum, P.~G., et al.\ 2003, \apjl, 587, L83
\bibitem[Vernet \& Cimatti(2001)]{2001A&A...380..409V} Vernet, J., \& Cimatti, A.\ 2001, \aap, 380, 409
\bibitem[Wandel et al.(1999)]{1999ApJ...526..579W} Wandel, A., Peterson, B.~M., \& Malkan, M.~A.\ 1999, \apj, 526, 579
\bibitem[Wang et al.(2004)]{2004ApJ...613..655W} Wang, W.-H., Cowie, L.~L., \& Barger, A.~J.\ 2004, \apj, 613, 655
\bibitem[Webb et al.(2003)]{2003ApJ...587...41W} Webb, T.~M., et al.\ 2003, \apj, 587, 41
\bibitem[Worsley et al.(2005)]{2005MNRAS.357.1281W} Worsley, M.~A., et al.\ 2005, \mnras, 357, 1281
\end{thebibliography}
\end{document}